\documentstyle[12pt,graphicx,epsf]{article}
\begin{document}

\def\AEF{A.E. Faraggi}

\def\vol#1#2#3{{\bf {#1}} ({#2}) {#3}}
\def\NPB#1#2#3{{\it Nucl.\ Phys.}\/ {\bf B#1} (#2) #3}
\def\PLB#1#2#3{{\it Phys.\ Lett.}\/ {\bf B#1} (#2) #3}
\def\PLA#1#2#3{{\it Phys.\ Lett.}\/ {\bf A#1} (#2) #3}
\def\PRD#1#2#3{{\it Phys.\ Rev.}\/ {\bf D#1} (#2) #3}
\def\PRL#1#2#3{{\it Phys.\ Rev.\ Lett.}\/ {\bf #1} (#2) #3}
\def\PRT#1#2#3{{\it Phys.\ Rep.}\/ {\bf#1} (#2) #3}
\def\MODA#1#2#3{{\it Mod.\ Phys.\ Lett.}\/ {\bf A#1} (#2) #3}
\def\RMP#1#2#3{{\it Rev.\ Mod.\ Phys.}\/ {\bf #1} (#2) #3}
\def\IJMP#1#2#3{{\it Int.\ J.\ Mod.\ Phys.}\/ {\bf A#1} (#2) #3}
\def\nuvc#1#2#3{{\it Nuovo Cimento}\/ {\bf #1A} (#2) #3}
\def\RPP#1#2#3{{\it Rept.\ Prog.\ Phys.}\/ {\bf #1} (#2) #3}
\def\APJ#1#2#3{{\it Astrophys.\ J.}\/ {\bf #1} (#2) #3}
\def\APP#1#2#3{{\it Astropart.\ Phys.}\/ {\bf #1} (#2) #3}
\def\EJP#1#2#3{{\it Eur.\ Phys.\ Jour.}\/ {\bf C#1} (#2) #3}
\def\etal{{\it et al\/}}

\newcommand{\cc}[2]{c{#1\atopwithdelims[]#2}}
\newcommand{\bev}{\begin{verbatim}}
\newcommand{\beq}{\begin{equation}}
\newcommand{\beqa}{\begin{eqnarray}}
\newcommand{\beqn}{\begin{eqnarray}}
\newcommand{\eeqn}{\end{eqnarray}}
\newcommand{\eeqa}{\end{eqnarray}}
\newcommand{\eeq}{\end{equation}}
\newcommand{\beqt}{\begin{equation*}}
\newcommand{\eeqt}{\end{equation*}}
\newcommand{\Eev}{\end{verbatim}}
\newcommand{\bec}{\begin{center}}
\newcommand{\eec}{\end{center}}
\def\ie{{\it i.e.}}
\def\eg{{\it e.g.}}
\def\half{{\textstyle{1\over 2}}}
\def\nicefrac#1#2{\hbox{${#1\over #2}$}}
\def\third{{\textstyle {1\over3}}}
\def\quarter{{\textstyle {1\over4}}}
\def\m{{\tt -}}
\def\mass{M_{l^+ l^-}}
\def\p{{\tt +}}

\def\slash#1{#1\hskip-6pt/\hskip6pt}
\def\slk{\slash{k}}
\def\GeV{\,{\rm GeV}}
\def\TeV{\,{\rm TeV}}
\def\y{\,{\rm y}}

\def\l{\langle}
\def\r{\rangle}
\def\LRS{LRS  }

\begin{titlepage}
\samepage{
\setcounter{page}{1}
\rightline{LTH--924} 
\vspace{1.5cm}
\begin{center}
 {\Large \bf OPERA data \\and\\ The Equivalence Postulate of Quantum Mechanics }
\vspace{.25 cm}

Alon E. Faraggi\footnote{
		                  E-mail address: faraggi@amtp.liv.ac.uk}
\\
\vspace{.25cm}
{\it Department of Mathematical Sciences\\
University of Liverpool, Liverpool, L69 7ZL, United Kingdom}
\end{center}

\begin{abstract}

An interpretation of the recent results reported by the OPERA collaboration 
is that neutrinos propagation in vacuum exceeds the speed of light.
It has been further been suggested that this interpretation can
be attributed to the variation of the particle average speed
arising from the Relativistic Quantum Hamilton Jacobi Equation.
I derive an expression for the quantum correction to the 
instantaneous relativistic velocity in the framework of the 
relativistic quantum Hamilton--Jacobi equation, which is derived 
from the equivalence postulate of quantum mechanics. While the 
quantum correction does indicate deviations from the 
classical energy--momentum relation, it does not necessarily 
lead to superluminal speeds. The quantum correction found herein
has a non--trivial dependence on the energy and mass of the particle, 
as well as on distance travelled. I speculate on other possible
observational consequences
of the equivalence postulate approach.

\end{abstract}
\smallskip}
\end{titlepage}

Recently the OPERA collaboration reported evidence for superluminal 
neutrino propagation from CERN to the Gran Sasso laboratory \cite{opera}. 
The arrival time of the muon neutrinos with average energy of $17{\rm GeV}$ 
is earlier by $\delta \equiv (v^2-1) =5\times 10^{-5}$ as compared to the 
speed of light in vacuum, and is reported with significance level of $6\sigma$.
The OPERA claim is is compatible with earlier results obtained by the MINOS 
experiment at FERMILAB, which measured the neutrino speed for energies
around $3{\rm GeV}$ and found \cite{minos} $\delta=(5.1\pm2.9)\times 10^{-5}$. 
These results are in an apparent conflict with the arrival time
of the supernova SN1987A that sets a limit of $\delta<2\times 10^{-9}$
for electron neutrinos with energies of the order of tens of MeVs \cite{sn1987a}.
 
If the OPERA results are confirmed it will indicate a departure from 
one of the pivotal tenants of fundamental physics. It will necessitate 
reexamination of the entire formulation of quantum field theories, which 
take the constancy of the speed of light and Lorentz invariance, as well
as causality as its basic assumptions. It is therefore an opportune moment 
to examine how deviations from the standard formalism may arise. 
The OPERA publication has indeed generated an avalanche of
papers that examine the experiment, the result and its potential 
consequences \cite{avalanche}. 

It has further been suggested that superluminal speeds are obtained
from a quantum version of the relativistic quantum Hamilton--Jacobi 
equation \cite{marco}. The quantum versions of the non--relativistic and
relativistic Hamilton--Jacobi equations have been derived from the equivalence
postulate of quantum mechanics \cite{fm,bfm}. The equivalence postulate is 
related to the existence of manifest phase--space duality \cite{fm,bfm},
which is also related to the classical--quantum duality proposed in \cite{fm1}. 
It has been shown that the equivalence postulate approach produces the
phenomenological characteristics of ordinary quantum mechanics, 
like tunnelling and energy quantisation for bound states \cite{fm}.  

In this paper I derive an expression for the quantum correction to the 
instantaneous relativistic velocity in the framework of the 
relativistic quantum Hamilton--Jacobi equation. While the 
quantum correction does indicate deviations from the 
classical energy--momentum relation, it does not necessarily 
lead to superluminal speeds. The quantum correction found herein
has a non--trivial dependence on the energy and mass of the particle, 
as well as on distance travelled.

I first examine the argument ref. \cite{marco}. The starting point is the Klein--Gordon
equation for a relativistic spinless free particle
\beq
(-\hbar^2 c^2 \Delta+m^2c^4-E^2)\psi=0 \ .
\label{kge}
\eeq
The Relativistic Stationary Quantum Hamilton--Jacobi equation follows 
by setting $$\psi=Re^{{1\over\hbar} S_0}.$$
\beq
(\nabla S_0)^2+m^2c^2-{E^2\over c^2}-\hbar^2{\Delta R\over R}= 0 \ ,
\label{rehje}
\eeq
where $S_0$ and $R$ satisfy the continuity equation
\beq
\nabla\cdot(R^2\nabla S_0)=0 \ .
\label{continuity}
\eeq
In terms of the quantum potential
\beq
Q=-{\hbar^2\over 2m}{\Delta R\over R} \ ,
\label{qp}
\eeq
and of the conjugate momentum
$p=\nabla S_0 \ ,$
the relativistic quantum Hamilton--Jacobi equation takes the form 
\beq
E^2=p^2c^2+m^2c^4+2m Qc^2 \ .
\label{e2p2}
\eeq
In one spatial dimension the continuity equation gives $$R={1\over{\sqrt{S_0^\prime}}},$$
and the quantum potential $Q$ takes the form
\beq
Q={\hbar^2\over4m}\{S_0,q\} \ ,
\label{qpin1dim}
\eeq
where $\{f,q\}={f'''\over f'}-{3\over2}\big({f''\over f'}\big)^2$ is the Schwarzian derivative of $f$.
Therefore
(\ref{rehje}) and (\ref{continuity}) reduce to the single equation
\beq
\left({{\partial S_0}\over{\partial q}}\right)^2+m^2c^2-{E^2\over c^2}+{\hbar^2\over2}\{S_0,q\}=0 \ .
\label{rhjein1dim}
\eeq
Equation (\ref{rhjein1dim}) is functionally similar to the nonrelativistic Quantum 
Hamilton--Jacobi equation. Hence, its solutions take the form
\beq
e^{{{2i}\over\hbar}S_0\{\delta\} }=e^{i\alpha}{{w+i{\bar\ell}}\over{w-i\ell}},
\label{s0solutions}
\eeq
where $w=\psi^D/\psi\in\ R$ and
$\psi$ and $\psi^D$ are two real linearly independent solutions of the
Klein-Gordon equation (\ref{kge}) in the 1+1 dimensional case.  Furthermore, we have
$\delta=\{\alpha,\ell\}$, with $\alpha\in\ R$ and
$\ell=\ell_1+i\ell_2$ integration constants. The necessary condition for the existence of 
a solution is that $\ell_1\ne 0$. This requirement is equivalent to having 
$S_0\ne cnst$,
which is a necessary condition to define the term $\{S_0,q\}$.

The crucial issue next is how to define the time evolution of the physical system. 
Floyd defines time parametrisation by using Jacobi's theorem \cite{floyd}
\beq
t-t_0={{\partial  S_0}\over{\partial  E}}=
{{\partial  S_0}\over{\partial  \psi}}{{\partial  \psi}\over{\partial  E}}+
{{\partial  S_0}\over{\partial  \psi^D}}{{\partial  \psi^D}\over{\partial  E}}+
{{\partial  S_0}\over{\partial  \ell}}{{\partial  \ell}\over{\partial  E}}+
{{\partial  S_0}\over{\partial {\bar\ell}}}{{\partial {\bar\ell}}\over{\partial  E}}
\label{jacobitheorem}
\eeq
where it is assumed that constants $\ell_1$ and $\ell_2$ may depend on $E$
as well \cite{fm,bfm}. Two linearly independent solutions of eq. (\ref{kge})
are given by 
\beq
\psi=\sin(kq)~~~\hbox{and}~~~~\psi^D=\cos(kq),
\label{psipsid}
\eeq 
where
$$
k={1\over{\hbar c}}\sqrt{E^2-m^2c^4}. 
$$
Ref. \cite{marco} then uses eq. (\ref{jacobitheorem}) to define a mean speed, 
which is given by 
 \beq
{q\over t}=
c{\sqrt{E^2-m^2c^4}\over E}
{
{\left\vert\sin(kq)-i\ell \cos(kq)\right\vert^2}
\over
{
\ell_1 +
{{\sin(2kq)}\over{2q}}
{{\partial \ell_1}\over{\partial k}}
-{{\sin^2(kq)}\over q}
\left(
\ell_1 {{\partial \ell_2}\over{\partial k}} -\ell_2 {{\partial \ell_1}\over{\partial k}}
\right)
}}
\label{meanspeed}
\eeq
This result differs slightly from the one derived in ref. \cite{marco}. 
The difference being in the last term that appears in the denominator, 
which is absent in ref. \cite{marco}. However, this discrepancy does not affect
the conclusions. From the form of eq. (\ref{meanspeed}) we can define the quantum correction
\beq
H_E\left(\ell_1, \ell_2;q\right) =
{
{\left\vert\sin(kq)-i\ell \cos(kq)\right\vert^2}
\over
{
\ell_1 +
{{\sin(2kq)}\over{2q}}
{{\partial \ell_1}\over{\partial k}}
-{{\sin^2(kq)}\over q}
\left(
\ell_1 {{\partial \ell_2}\over{\partial k}} -\ell_2 {{\partial \ell_1}\over{\partial k}}
\right)
}}
\label{qcorrection}
\eeq
to the classical relationship
\beq
v=c{\sqrt{E^2-m^2c^4}\over E}.
\label{classicalv}
\eeq
It is then clear that the classical limit $\hbar\rightarrow 0$ corresponds
to 
\beq
H_E\left(1, 0;q\right)=1.
\label{heeq1}
\eeq
It is then stated in \cite{marco} that in general 
\beq
H_E(\ell_1,\ell_2;q)>1.
\label{hegt1}
\eeq
However, this is obviously not the case. To examine the behaviour of $H_E$
we can study the case $\ell_1$ and $\ell_2$ are independent of $E$. 
In this case the last two terms in the denominator of $H_E$ are zero. 
The expression for the mean speed, eq. (\ref{meanspeed})
reduces to
\beq
{q\over t}=
c{\sqrt{E^2-m^2c^4}\over E}
{
{\left(\sin^2(kq)+(\ell_1^2+\ell_2^2)\cos^2(kq)+ \ell_2 \sin(2kq) \right)}
\over
{
\ell_1
}}.
\label{meanspeed2}
\eeq
To simplify this expression further we can examine the case $\vert\ell\vert=1$, 
with $\ell_1=\cos\alpha$, $\ell_2= \sin\alpha=$, and $\alpha={\rm constant}\ne0$. 
In this case the relativistic speed is multiplied by the factor
\beq 
v=v_{\rm rel}
{
{\left(1+ \sin\alpha \sin(2kq) \right)}
\over
{
\cos\alpha
}}.
\label{simplifiedfactor}
\eeq
Since $q$ is some arbitrary distance, we can take $q=\pi/(4k)$ in which case the
quantum correction eq. (\ref{qcorrection}) reduces
\beq
{{1+\sin\alpha}\over{\cos\alpha}}.
\label{simpleqc}
\eeq 
Since a priori there is no reason to restrict $\alpha$ we 
see that in general the quantum factor $H_E(\ell_1,\ell_2;q)$ 
is not larger than 1. Hence, the quantum correction
does not indicate, in general, the existence of superluminal 
motion. 

Nevertheless, OPERA data may indeed indicate deviations from the 
relativistic energy--momentum relation eq. (\ref{classicalv}). To 
study this question we can use the Jacobi theorem, eq. 
(\ref{jacobitheorem}) to define the instantaneous speed
in the quantum case. For this purpose we can 
use eq. (\ref{rhjein1dim}) to rewrite 
eq. (\ref{jacobitheorem}) in the form 
\beq
t-t_0={\partial\over\partial E}\int^q_{q_0}dx
{{\partial S_0}\over{\partial x}}=
\int^q_{q_0} dx 
{
{E/c^2-mc^2\partial_EQ}
\over
{\left(E^2/c^2-m^2c^2-2mc^2Q\right)^{1\over2}}
}
\label{jtint}
\eeq
where $q_0=q(t_0)$. The velocity is given by
\beq 
{{d q}\over{d t}} = \left( {{dt}\over {dq}}\right)^{-1}= 
{{\partial_q S_0} \over {E/c^2-m c^2\partial_E Q}}
\label{instanteneousv}
\eeq
Hence, we have that the instantaneous velocity is 
given by
\beq
{\dot q}= 
{
p
\over
{\left(E/c^2-mc^2\partial_E Q \right)
}}
=
{
{pc^2}
\over 
{E\left(1-{m\over E}{{\partial Q}\over {\partial E}}c^4\right)
}}
\label{qdot}
\eeq 
Making the approximation  
$$
{m\over E}{{\partial Q}\over {\partial E}}c^4 << 1, 
$$
we have 
\beq
{\dot q}= {p\over E}c^2\left(1+ {m\over E}{{\partial Q}\over {\partial E}}c^4\right). 
\label{approxqdot}
\eeq
Hence, we see that in this case the classical relativistic relation
$$
{\dot q}={p\over E}c^2 
$$
is modified by the quantum factor
\beq 
H_E(\ell_1, \ell_2;q) = \left(1 +  {m\over E}{{\partial Q}\over {\partial E}}c^4\right),
\label{poeqf}
\eeq
which vanishes in the classical limit $\hbar \rightarrow 0\Rightarrow Q\rightarrow 0$, 
and where $Q$ is given by eq. (\ref{qp}) in the higher dimensional case and by
eq. (\ref{qpin1dim}) in the one dimensional case. 

A priori we have no reason to assume that the quantum correction term in eq. (\ref{approxqdot}),
{\it i.e.}
$$
 {m\over E}{{\partial Q}\over {\partial E}}c^4, 
$$
is positive definite. In particular we have no reason to assume that the slope 
of the quantum potential $Q$ with respect to the energy $E$ is positive definite. 
Hence, we have no reason to infer that the quantum correction factor 
(\ref{poeqf}) is larger than 1. By using the expression given in (\ref{qpin1dim})
we can study this question in the case of the
relativistic stationary quantum Hamilton-Jacobi equation.
Using eqs. (\ref{s0solutions}, \ref{psipsid}) and the expression for $Q(q)$ 
given in eq. (\ref{qpin1dim}) we obtain
\beqn
{{4m}\over \hbar^2} Q(q) & = &
{k^2\over 
{4\left(\cos^2(kq) + (\ell_1^2+\ell_2^2)\sin^2(kq) + \ell_2\sin(2kq)\right)^2} 
} \cdot  \nonumber
\\
& &
\left(3-6\ell_1^2+3\ell_1^4+6\ell_2^2+6\ell_1^2\ell_2^2+3\ell_2^4 \right. \nonumber\\
& & -4 (-1+\ell_1^4+2 \ell_1^2\ell_2^2+\ell_2^4)\cos(2kq)\nonumber\\
& & +( 1+\ell_1^4-6\ell_2^2+\ell_2^4+2\ell_1^2(-1+\ell_2^2))\cos(4kq)\nonumber\\
& & + 8 \ell_2\sin(2kq) + 8 \ell_1^2\ell_2\sin(2kq)+ 8 \ell_2^3 \sin(2kq)\nonumber\\
& & \left. +4\ell_2\sin(4kq) -4\ell_1^2\ell_2\sin(4kq) -4\ell_2^3\sin(4kq) \right)
\eeqn 
Similar to the analysis in the case of eqs. (\ref{meanspeed})
we can examine the corrections to the classical relationship in 
special cases. We first note that the case with $\ell_1=1$ and $\ell_2=0$
we have that $Q\equiv0$. Therefore in this case the classical relation is
not affected. This is in agreement with the result found in eq. (\ref{heeq1}) \cite{marco},
which showed that this choice of the $\ell_{1,2}$ parameters reproduces the classical 
relativistic result. We note that this is in contradiction to the requirement 
that $Q(q)\ne 0$ always, which is a necessary consequence of the equivalence postulate. 
We conclude that $(\ell_1, \ell_2)=(1,0)$ is not an allowed point in the parameter space. 

We can further examine the behaviour of the quantum correction to the classical result
by taking other limiting cases, {\it i.e.} setting $\ell_1=\cos\alpha$ and $\ell_2=\sin\alpha$. 
The result is not very illuminating and we can simplify it further by setting $\alpha=\pi/4$. 
Since the partial derivative $\partial_E k$ is positive definite, we need only focus
on the partial derivative $\partial_k Q$. In the special case examined here we obtain
\beq
{{\partial Q}\over{\partial k}}=
{
{
(2k (20+8\sqrt{2}kq\cos(2kq)-12 \cos(4kq)+23\sqrt{2}\sin(2kq)-\sqrt{2}\sin(6kq)))
}
\over
{
(2+\sqrt{2}\sin(2kq))^3
}
}.
\label{pqpk}
\eeq
Setting $q=(2\pi)/(4k)$ gives $\partial_k Q\approx -2.4k. $ It is noted that similar to the case of
eq. (\ref{meanspeed}) the quantum correction is not necessarily larger than 1, though
larger than 1 factors are clearly possible.
Examining the results obtained both in eq. (\ref{meanspeed}) and (\ref{approxqdot})
we conclude the quantum correction to the energy--momentum 
relation does lead to deviations from the classical result. However, the 
correction does not necessarily lead to superluminal speeds. 
The quantum contribution has a complicated dependence on
the $\ell_1$ and $\ell_2$ parameters, as well as a nontrivial 
dependence on the energy and distance travelled through the
$k$ and $q$ variables, respectively. Furthermore, as seen in eq.
(\ref{approxqdot}) there is a flavour dependence that arises from the 
particle mass. 

The OPERA data, if confirmed by future experiments, may indicate a paradigm shift
from the established physics foundations, perhaps discerning between
the equivalence postulate approach and the conventional approaches to quantum
mechanics. In this respect it is also of interest to note that energy quantisation 
arises in the equivalence postulate approach due to the consistency 
requirement that the wave--function is continuous on the extended real line \cite{fm}. 
This requirement is reminiscent of quantisation in compact spaces. We may therefore
speculate that the equivalence postulate approach implies that the universe has a compact 
topology. Investigation of this question has for example been recently discussed in \cite{am}.
Observations of large angle correlations in the cosmic  microwave background radiation
in contemporary experiments may therefore lend support to the equivalence postulate 
approach to quantum mechanics. There are two key ingredients of this approach \cite{fm,bfm}.
The first is a quadratic identity which is a manifestation of the quantum Hamilton--Jacobi
equation in its non--relativistic or relativistic forms. The second is 
a co--cycle condition which manifests the symmetry properties of the formalism. 
In one dimension it is invariant under Mobi\"us transformations of the coordinate $q$, which 
in one dimension uniquely fixes the quantum potential to be given by the Schwarzian 
derivative, eq. (\ref{qpin1dim}). In higher Euclidean dimensions the cocycle condition is invariant
under $D$--dimensional Mobi\"us transformations, whereas it is invariant with respect to 
the $D+1$ conformal group in the case of Minkowski space \cite{bfm}. This may indicate
the relevance of the conformal approach to gravity \cite{conformalgravity}.
The conformal approach to quantum gravity generically suffers from the existence of
ghosts. Ref. \cite{maldacena}, however, argued that in the presence of certain boundary 
conditions the ghosts are removed and hence enabling a consistent formulation of the
theory. The equivalence postulate approach to quantum mechanics, which manifests
the relevance of the conformal group in $D+1$ Minkowski space, may provide a viable
framework to formulate quantum gravity. OPERA data may have thus opened the 
door to embark on that journey. 

\section*{Acknowledgements}

I would like to heartily thank Marco Matone for our collaboration
over the years and for the warm hospitality extended to me and my 
family on numerous visits to Schio.
This work was supported in part by the STFC (PP/D000416/1).

\end{document}